\documentclass[11pt,a4paper]{article}
\usepackage[a4paper,
left=2.5cm,
right=2.5cm,
top=2.5cm,
bottom=2.5cm]{geometry}

\usepackage{lmodern}
\usepackage[colorlinks,hypertexnames=false]{hyperref}
\usepackage{caption}
\usepackage{cite}

\usepackage{amsmath}
\usepackage{amssymb}
\usepackage{amsfonts}
\usepackage{amsthm}
\usepackage{bm}
\usepackage{mathrsfs}

\usepackage{graphicx}
\usepackage{booktabs}
\usepackage{multirow}
\usepackage{array}

\usepackage{hyperref}

\begin{document}

%%%%%%%%%%%%%%%%%%%%%%%%%%%%%%%%%%%%%%%%%%%%%%%%%%%%%%%%%%%%%%%%%%%%%%
%%%%%%%%%%%%%%%%%%%%%%%%%%%% Title %%%%%%%%%%%%%%%%%%%%%%%%%%%%%%%%%%%
%%%%%%%%%%%%%%%%%%%%%%%%%%%%%%%%%%%%%%%%%%%%%%%%%%%%%%%%%%%%%%%%%%%%%%

\begin{center}

{\LARGE\bfseries
Vortex State of Ultralight Dark Matter and the Fornax Timing Problem
\par}

\vspace{0.8cm}

{\large
V.~Gorkavenko$^{1,2}$,
O.~Barabash$^{1}$,
T.~Gorkavenko$^{1}$,
K.~Korshynska$^{3,4}$,
O.~Teslyk$^{1}$,
A.~Zaporozhchenko$^{1}$,
E.~Gorbar$^{1,2}$
}

\vspace{6mm}

{\it
$^{1}$Faculty of Physics,
Taras Shevchenko National University of Kyiv,\\
64/13 Volodymyrska Street,
Kyiv 01601,
Ukraine
}

\vspace{2mm}

{\it
$^{2}$Bogolyubov Institute for Theoretical Physics,
National Academy of Sciences of Ukraine,\\
14-b Metrolohichna Street,
Kyiv 03143,
Ukraine
}

\vspace{2mm}

{\it
$^{3}$Institut f\"ur Mathematische Physik,
Technische Universit\"at Braunschweig,\\
Mendelssohnstra\ss e 3,
38106 Braunschweig,
Germany
}

\vspace{2mm}

{\it
$^{4}$Fundamentale Physik f\"ur Metrologie (FPM),
Physikalisch-Technische Bundesanstalt (PTB),\\
Bundesallee 100,
38116 Braunschweig,
Germany
}

\end{center}

%%%%%%%%%%%%%%%%%%%%%%%%%%%%%%%%%%%%%%%%%%%%%%%%%%%%%%%%%%%%%%%%%%%%%%

\begin{abstract}
We investigate the impact of the vortex state of the ultralight dark matter (ULDM) on the dynamical friction acting on moving globular clusters. 
Comparing this force with that for the solitonic ground state, it is shown that the internal structure and rotation of the ULDM core strongly affect the orbital decay of globular clusters. In particular, co-directional rotation in a vortex state can lead to significant suppression of dynamic friction  {at certain distances where globular clusters and ULDM velocities match}. Applying these findings to the Fornax dwarf galaxy, it is found that the Fornax timing problem is naturally alleviated.

\medskip

\noindent\textbf{Keywords:}
ultralight dark matter, globular clusters, Fornax timing problem, dynamical friction force

\end{abstract}

\section{Introduction}

The dynamics of satellite galaxies and their internal stellar systems provide a powerful probe of the fundamental properties of dark matter and its interaction with baryonic matter. In particular, the orbital evolution of massive substructures, including globular clusters within dark-matter-dominated systems, offers a sensitive test of gravitational dynamics on galactic and subgalactic scales. One of the most intriguing and well-studied examples in this context is the Fornax dwarf spheroidal galaxy \cite{Pace}, a satellite of the Milky Way that hosts several globular clusters at relatively large galactocentric distances which exist more than 10 Gyr.

Within the standard cold dark matter (CDM) paradigm, globular clusters orbiting inside a massive dark matter halo are expected to experience substantial dynamical friction due to gravitational interactions with background dark matter particles and baryons. This frictional force should lead to orbital decay, causing the clusters to spiral toward the galactic center on timescales that, for some of the observed globular clusters in Fornax, are significantly shorter than the age of the Universe. This apparent contradiction between theoretical predictions and observations is commonly referred to as the Fornax timing problem \cite{Oh,Meadows,Bar:2021jff}.

The persistence of this discrepancy has motivated extensive exploration of alternative explanations, including modifications of the dark matter model itself. In this regard, ultralight dark matter (ULDM) has emerged as a particularly compelling candidate \cite{Marsh:2015xka,Lee:2017qve,Ferreira:2020fam,Matos:2023usa,Schive:2025bcm}. ULDM consists of extremely light bosonic particles whose de Broglie wavelength is comparable to galactic scales, leading to pronounced wave-like behavior on astrophysical distances. Due to the extremely large de Broglie wavelength, of the order of kiloparsecs, ULDM models naturally suppress the formation of small-scale structures, in contrast to CDM scenarios, which tend to overpredict the abundance of dwarf galaxies and the amount of dark matter in galactic centers.

These wave effects can suppress dynamical friction and modify the orbital evolution of massive objects embedded in ULDM halos. In our previous work \cite{Gorbar:2025rrd}, we investigated the Fornax timing problem within the ULDM scenario by assuming a spherically symmetric ground-state configuration of a Bose–Einstein condensate soliton. By incorporating a damping term in the generalized Gross–Pitaevskii equation, we analyzed the time evolution of globular cluster orbits and demonstrated that the solitonic structure of the ULDM halo can substantially affect their dynamical evolution. Analytical formulas for the dynamical friction force were used to contrast the infall times and dynamical evolution of globular clusters in the presence and absence of the damping term.

In the present paper, we extend the analysis of the dynamical friction force acting on a moving massive object, such as a globular cluster, by considering a rotating vortex state of ultralight dark matter. We study how the internal rotation and angular momentum of the ultralight dark matter core influence the dynamical friction force \cite{Chandrasekhar,Bondi,Dokuchaev,Ruderman,Rephaeli,Ostriker} exerted on globular clusters. In particular, we focus on determining the characteristic timescales over which the velocities and orbits of globular clusters undergo significant changes, and we assess the implications of vortex-induced effects for the long-term stability of globular cluster systems in dwarf galaxies such as Fornax.

  The paper is organized as follows. The ground and vortex states of the ULDM soliton are described in Sec.2.  Numerical analysis of the characteristic time of the velocity change of globular clusters in the Fornax dwarf galaxy is given in Sec.3. The results are summarized in Sec.4.

\section{Ground and vortex states of ULDM soliton}

Ultralight dark matter can be accurately described as a complex classical field $\psi$ forming a Bose–Einstein condensate (BEC) on galactic scales \cite{Ferreira:2020fam}. Owing to the extremely small mass of ULDM particles, their de Broglie wavelength is comparable to galactic scales, leading to macroscopic quantum coherence and wave-like behavior of dark matter halos. In this regime, the collective dynamics of the ULDM field cannot be captured by a particle description and must instead be treated in view of large occupation number by using a classical field approach.
The dynamical evolution of the self-gravitating and self-interacting BEC field $\psi$, together with its self-consistent gravitational potential $\Phi$, is governed by the coupled Gross–Pitaevskii–Poisson (GPP) system of equations \cite{Ferreira:2020fam,Calmet:2015fua}
\begin{equation}
i\hbar\frac{\partial\psi}{\partial t} = \left(-\frac{\hbar^{2}}{2m}\nabla^{2} + gN|\psi|^{2} + m\Phi \right)\psi,
\label{GPP 1}
\end{equation}
\begin{equation}
\nabla^{2}\Phi = 4\pi GmN|\psi|^{2},
\label{GPP 2}
\end{equation}
where $m$ is the mass of the ULDM particle, $N$ is the total number of bosons, $\hbar$ is the Planck constant, and $G$ is the gravitational constant. The nonlinear coupling
$g = {4 \pi \hbar^{2}a_{s}}/{m}$ 
characterizes the short-range self-interaction between ULDM particles, with $a_s$ being the $s$-wave scattering length.
 
Repulsive self-interaction of dark matter particles corresponds to $g>0$, while $g<0$ describes the attractive case. The last term in Eq.~(\ref{GPP 1}) represents the gravitational interaction mediated by the potential $\Phi$, which is sourced by the mass density of the condensate via Eq.~(\ref{GPP 2}) and is responsible for the formation of localized solitonic cores.
Approximate analytical, variational, and numerical solutions of the GPP equations have been extensively studied in the literature, revealing the existence of stable solitonic ground states and excited configurations \cite{Nikolaieva:2021owc,Korshynska:2023kxa}. 

In the present paper, we restrict our attention to the regime of weak repulsive self-interaction between ULDM bosons and adopt a Gaussian variational ansatz \cite{chavanis2011mass} for the density profile of the spherically symmetric soliton ground state,
\begin{equation}\label{rhoGR}
\rho_0 (r) = \rho_c e^{-r^2/R^2},
\end{equation}
where $\rho_c$ is the central density and $R$ characterizes the spatial extent of the solitonic core.
To describe rotating configurations of the condensate, we consider a toroidal rotated vortex state aligned along the $z$-axis, whose density profile is given by
\begin{equation}\label{rho1}
\rho_1(\mathbf{r}) = \rho_c \frac{r_\perp^2 e^{-r^2/R^2}}{r_\perp^2 + (2\xi)^2},
\end{equation}
where $r_\perp = \sqrt{x^2 + y^2}$ is the cylindrical radial coordinate and $\xi$ denotes the coherence length, which sets the characteristic size of the vortex core, see \cite{Nikolaieva:2021owc} for details. For the Fornax dwarf spheroidal galaxy, we adopt the fiducial parameters $R = 581~\mathrm{pc}$ and $\rho_c = 5.31 \times 10^{-2}~M_\odot/\mathrm{pc}^3$, consistent with observational constraints.
The coherence length $\xi$ is determined by the balance between the kinetic and interaction energies of the condensate and can be expressed as
\begin{equation}
\xi = 6.49 \times 10^{-22} \left(\frac{1~\mathrm{eV}/c^2}{m} \frac{1~\mathrm{fm}}{a_s}\right)^{1/4}~\mathrm{kpc},
\label{eq: coherence length}
\end{equation}
highlighting the sensitivity of coherence length to both the particle mass and the strength of self-interaction.

 {Considering a more accurate model of the Fornax galaxy, we have to take into account the baryonic matter with density $\rho_B(r)$, which produces the gravitational potential $\Phi_B (r)$. In dwarf galaxies, the baryonic density can be modeled by the Plummer sphere \cite{plummer1911problem}, which reads
\begin{equation}
    \rho_B(r) = \frac{3 b^2 M_B}{4 \pi (r^2 + b^2)^{5/2}} \label{eq: rhob}\\
\end{equation}
and is characterized by the total baryon mass $M_B$ and the Plummer radius $b$. For Fornax these parameters are $M_B = 3\times 10^7 M_\odot$ and $b = 668$~pc \cite{cappellari2006sauron, kowalczyk2019schwarzschild}. Measuring mass in the units of total DM mass $M_{DM}$ and distance in the units of DM size $R$, we obtain $M_B/M_{DM} = 0.52$ and $b/R = 1.15$. The gravitational potential and the rotation curve can be easily obtained in this case
\begin{eqnarray}
   & \Phi_B(r) = - \frac{G M_B}{\sqrt{r^2 + b^2}} \, ,\label{eq: phib}\\
    & v^2_B = r\frac{d \Phi_B}{dr}=GM_B\,\frac{r}{\left(r^{2}+b^{2}\right)^{3/2}}\,. \label{eq: vb}
\end{eqnarray}
}

For the analytically determined Gaussian density of ultralight dark matter \eqref{rhoGR},  {the gravitational potential (solution of the Poisson equation) is given by
\begin{equation}
   \Phi_{0}(r) = -G M_{DM} \frac{\textrm{Erf}(r/R)}{r},
\end{equation} 
where $\textrm{Erf}(x)$ denotes the error function.} One can derive the rotation curve for the orbiting objects in the spherically symmetric soliton state of ULDM \cite{chavanis2019mass}  {and the baryonic matter}
\begin{equation}\label{v2}
    v^2_{0B}(r)  {=r\frac{d (\Phi_{0}+\Phi_{B})}{dr}}=\frac{GM_{DM}}{R} \left[\frac{R}{r} \textrm{erf} \left(\frac{r}{R}\right) - \frac{2}{\sqrt{\pi}}e^{-(r/R)^2}\right]+ {v^2_B}.
\end{equation}
% that can be written in a more convenient form
% \begin{equation}\label{v2_1}
%     \frac{v(r)}{1 \textrm{km}/\textrm{s}} = 20.7 \times \left[ \frac{581~\textrm{pc}}{r} \textrm{erf} \left(\frac{r}{581~\textrm{pc}}\right) - \frac{2}{\sqrt{\pi}}\exp \left[-\left(\frac{r}{581~\textrm{pc}}\right)^2\right] \right]^{1/2}.
% \end{equation}

\begin{figure}[t]
    \centering  \includegraphics[width=\textwidth]{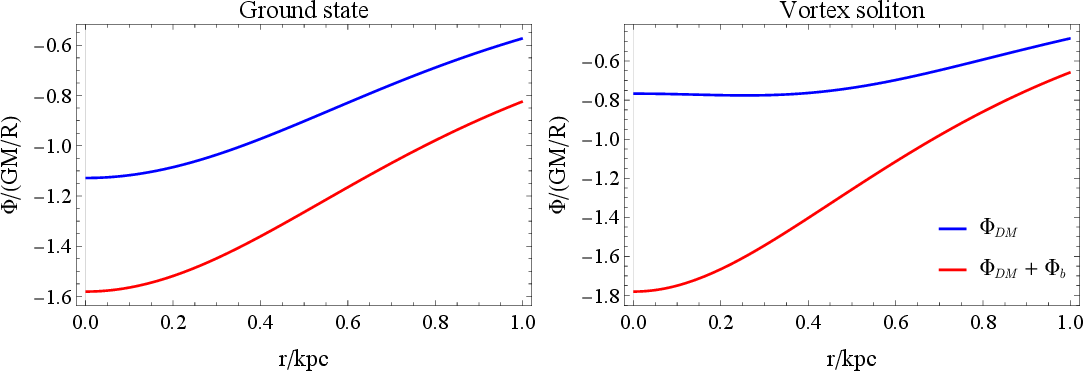}
    \caption{ {The gravitational potential with and without baryonic matter in ground (left) and vortex (right) cases. We take  $\xi/R = 0.5$ for the vortex state.}}\label{figure1}
\end{figure}

In the case of ULDM  in the form of a rotating vortex soliton,  {the gravitational potential at a distance $\mathbf{r}$ from the centre of the galaxy in the plane $z=0$ is given by
\begin{multline}
    \Phi_{vortex}(\mathbf{r}) = - G \int d\mathbf{r}' \frac{\rho_1(\mathbf{r}')}{|\mathbf{r} - \mathbf{r}'|}=\\= -  G \int_0^{+\infty} dr' (r')^2 \int_0^\pi d \theta' \sin \theta' \int_0^{2\pi} d\phi'  \frac{\rho(\mathbf{r}')}{\sqrt{r^2 + (r')^2 - 2rr'\sin \theta' \cos \phi'}}.
\end{multline}
For the vortex soliton case, the vanishing of the DM density in the center of the soliton creates a local maximum at the origin, and thus for the vortex soliton $\partial \Phi_{vortex}/\partial r < 0$ in some region $r \in [0, r_\textrm{crit}]$. This implies that no stable rotation curves exist in the region $r < r_\textrm{crit}$. To ensure the existence of stable rotation curves, it is necessary to take into account the baryonic matter contribution. We present the gravitational potential for the ground and vortex soliton states with and without the contribution of the baryonic matter in Fig.\ref{figure1}. It should be noted that the minimum of the potential is shifted from $r/R = r_\textrm{crit}/R = 0.445$ to $r = 0$, when the baryonic contribution is added. After taking the baryon matter contribution, the rotation curve becomes well defined
\begin{equation}
    v^2_{1B}(r)=r\frac{d \Phi_{vortex}}{dr}+v^2_B,
\end{equation}
where $r \times d \Phi_{vortex}/dr$ should be taken numerically.
}

% the rotation curves differ significantly only in the region close to the center of the galaxy. 
% Since globular clusters move rather far from the galactic center, we do not provide the corresponding formula for the rotation curves in this case.

The vortex solution is described by a wave function whose angular dependence is given by $\psi_1 \sim e^{i\,s\,\phi}$ ($s=1$), i.e., it carries quantized angular momentum, which gives rise to a nonzero particle current
\begin{align}
\mathbf{j} = -\frac{i\hbar}{2m}(\psi_{1}^{} \boldsymbol{\nabla}\psi_{1} - \psi_{1} \boldsymbol{\nabla}\psi_{1}^{})
= \frac{\hbar}{m}\frac{|\psi_{1}|^{2}}{r_\perp}\mathbf{e}_{\phi}.
\end{align}

This current defines the velocity field $\mathbf{u}(\mathbf{r})$ of ULDM particles circulating around the vortex axis
\begin{equation}\label{u}
    \mathbf{u}(\mathbf{r})=\frac{\frac{\hbar}{m}\frac{|\psi_{1}|^{2}}{r_\perp}}{|\psi_{1}|^{2}}\, \mathbf{e}_{\phi}.
\end{equation}
In physical units, the corresponding velocity distribution reads
\begin{equation}
\frac{\mathbf{u}}{1~\mathrm{m}/\mathrm{s}} =
1.92 \times 10^{-15} \frac{1~\mathrm{eV}/c^2}{m} \frac{1~\mathrm{pc}}{r_\perp}\mathbf{e}_{\phi},
\label{velocity-distribution}
\end{equation}
demonstrating that the rotational motion is most pronounced near the vortex core and decreases with radial distance.

Number $s$ defining the angular dependence of the wave function $\psi$ is the so-called topological charge, which determines the angular momentum of the ULDM state. The state with $s=0$ describes the spherically symmetric ULDM soliton ground state, while the state with the wave function $\psi_1$ and $s=1$ describes the toroidal ULDM soliton rotating around the center of the galaxy. States with higher topological charge $s>1$ are unstable \cite{Dmitriev:2021utv,Nikolaieva:2021owc}.

By minimizing the total energy of the solitonic configuration, one obtains the corresponding mass–radius relation for the ULDM condensate \cite{chavanis2011mass}
\begin{equation}
R = \frac{\sigma}{\nu} \frac{\hbar^2}{G M_{DM}m^2} \left(1 \pm \sqrt{1 + \frac{6 \pi \zeta \nu}{\sigma^2} \frac{Gm M^2a_s}{\hbar^2}}\right),
\label{eq: first mass-radius relation}
\end{equation}
where $M_{DM}$ is the total mass of the soliton DM and the numerical coefficients are given by $\sigma = 3/4$, $\nu = 1/\sqrt{2 \pi}$, and $\zeta = 1/(2 \pi)^{3/2}$. The plus sign corresponds to repulsive or vanishing self-interaction ($a_s \geq 0$), while the minus sign applies to the attractive case ($a_s<0$). In the following, we focus exclusively on the $a_s \geq 0$ regime. Note the presence of explicit dependence on the Planck constant, which explicitly demonstrates the quantum-mechanical nature of this relation.

Substituting the observationally inferred parameters for the Fornax dwarf galaxy, namely, its radius $R$ and the total DM mass $M_{DM} = 5.8 \times 10^7 M_\odot$ \cite{jardel2012dark}, into Eq.~(\ref{eq: first mass-radius relation}), one obtains the mass-radius relation between the ULDM particle mass $m$ and the scattering length $a_s$ given by
\begin{equation}
\left(\frac{m}{\mathrm{eV}/c^2} \right)^2 =
4.76 \times 10^{-44} \left(1 + \sqrt{1 + 1.21 \times 10^{83} \frac{a_s}{\mathrm{fm}} \frac{m}{\mathrm{eV}}}\right).
\label{eq: first mass-radius relation2}
\end{equation}
In the non-interacting limit $a_s = 0$, this expression yields the characteristic ULDM particle mass $m = 3.09 \times 10^{-22}~\mathrm{eV}$. For larger particle masses, a nonzero repulsive self-interaction becomes necessary.
For smaller values of dark matter particle masses, it follows from Eq.~(\ref{eq: first mass-radius relation}) that the repulsion between dark matter particles should be replaced by attractive interaction.

 {Relation \eqref{eq: first mass-radius relation2} allows us to uniquely express the coupling of self-interaction ($g = {4 \pi \hbar^{2}a_{s}}/{m}$) between dark matter particles in terms of their mass and consider the characteristic time only as a function of the
DM particles' mass and the state of dark matter (ground or vortex).}

\section{Characteristic time of the velocity change}

Having established the theoretical framework for dynamical friction in ultralight dark matter and outlined the method for estimating the characteristic timescale of orbital evolution, we now proceed to a quantitative analysis. In this section, we present numerical results for the dynamical friction acting on globular clusters in both the solitonic ground state and the rotating vortex configuration of the ULDM halo. We focus on how the characteristic timescale depends on the cluster mass, its orbital radius, and the relative orientation between the cluster motion and the ULDM flow, and discuss the implications of these results for the Fornax timing problem.

To perform a qualitative analysis of the impact of dynamical friction on the motion of a globular cluster, it is sufficient to approximate the globular cluster as a point-like object and neglect corrections associated with its finite spatial extent and internal structure \cite{Gorkavenko:2024upe,Barabash:2025ylw}.  This approximation is well justified as long as the characteristic size of the cluster is much smaller than the typical length scales over which the dark matter density and velocity field vary. In this regime, the dominant contribution to dynamical friction arises from the bulk gravitational interaction between the moving cluster and the surrounding dark matter background. In particular, when the Plummer radius is smaller than about $10^{-2}$ of the radius of its orbit, the dynamical friction force is essentially indistinguishable from that of a point-like object.

For a point-like probe moving with a constant velocity along a circular orbit, the dynamical friction force in an ultralight dark matter medium was calculated in \cite{Desjacques_2022,Buehler:2022tmr}.  In the framework of linear response theory, the backreaction of the medium on the moving object leads to the formation of a gravitational wake, which in turn exerts a drag force opposing the motion of the probe.

A convenient way to characterize the cumulative effect of this force, which we adopt in our study here is through the characteristic timescale over which the probe experiences a substantial change in its velocity \cite{Hui:2016ltb},
\begin{equation}\label{probeT}
T = \frac{ {v_{GC}} M}{F_{ {fr,ULDM}}+ {F_{fr,B}}},
\end{equation}
where  {$v_{GC}$ denotes the velocity of the probe moving in the plane $z=0$}, 
%and the local ULDM flow with velocity $\mathbf{u}$, 
 $M$ is the mass of the probe, $F_{fr,ULDM}$ is the tangential component of the dynamical friction force acting on the probe   {in ULDM environment},   {and $F_{fr,B}$ is dynamical friction force acting on probe in the baryonic matter environment.}
 
The explicit expression for the dynamical friction force  {acting on probe in ULDM environment} at its position $\vec r = \vec r_p(t)$ reads \cite{Berezhiani:2023vlo}
\begin{equation}\label{Fforce}
\mathbf{F}_{fr,ULDM}(t)=\frac{G^2 M^2 \rho_\mathrm{DM}}{\pi^2}
\!\!\int\limits_{0}^{+\infty}\!\! d\tau
\!\!\int\!\! d\omega
\!\!\int\!\! d\mathbf{k}
\frac{e^{-i \omega \tau + i \vec k \cdot (\vec r_p(t)-\vec r_p(t-\tau))}}
{c_s^2 \vec{k}^2 + \frac{(\vec{k}^2)^2}{4m^2} - (\omega - i\varepsilon)^2},
\end{equation}
where $\rho_\mathrm{DM}$ is the dark matter density evaluated at the location of the probe. The denominator encodes the dispersion relation of density perturbations in ULDM, with the first term corresponding to the effective pressure of the condensate, while the second term arises from the quantum pressure associated with the wave nature of the ultralight dark matter. 
The square of adiabatic sound speed $c_s$ is defined as derivative of pressure of ULDM with respect to density. The adiabatic sound speed is given by 
\begin{equation}
\label{cs}
c_s = \sqrt{\frac{\rho_\mathrm{DM} g}{m^2}},
\end{equation}
and depends on both the local dark matter density and the strength of the self-interaction.  This parameter plays a crucial role in determining the efficiency of wake formation and, consequently, the magnitude of the dynamical friction force.

For a test particle of mass $M$ (star or a globular cluster of the corresponding mass), moving with a constant orbital velocity $v$ along a circular trajectory of radius $r$ inside a  {ultralight} dark matter  {soliton} core, the tangential component of the dynamical friction force can be written in a compact analytic form. The friction force \eqref{Fforce} acting opposite to the direction of motion  can be conveniently presented in the form
\begin{equation}\label{forcekhuriFull_re}
F_{\rm fr}=\frac{4\pi G^2 {M}^2 \rho_{\rm DM}}{c_s^2}\mathcal{F},
\end{equation}
where the dimensionless factor $\mathcal{F}$ encodes the full dependence of the dynamical friction force on the kinematic and microscopic properties of the system. In particular, it accounts for the relative velocity between the star and the dark matter background, as well as for the wave-like response of the superfluid medium.
Following the results obtained in Refs.~\cite{Desjacques_2022,Buehler:2022tmr,Berezhiani:2023vlo}, the  dimensionless friction force $\mathcal{F}$ can be represented as a double sum over angular momentum quantum numbers,
\begin{equation}
\mathcal{F}= \sum_{\ell=1}^{\ell_{\rm max}}\sum_{m_l=-\ell}^{\ell-2}\gamma_{\ell m_l}\,
\text{Im}\!\left(S_{\ell,\ell-1}^{m_l}-{S_{\ell,\ell-1}^{m_l+1}}^{*}\right),
\label{FDF1_re}
\end{equation}
where $\ell$ and $m_l$ play the role of azimuthal and magnetic quantum numbers, respectively. This decomposition reflects the fact that the gravitational wake generated by the moving perturber can be expanded in spherical harmonics, in close analogy with the classic treatment of dynamical friction in a gaseous medium. Indeed, the derivation of this expression closely parallels the logic of Ostriker’s formula for a massive object moving through a homogeneous fluid \cite{Ostriker}, but with important modifications arising from the quantum and superfluid nature of ultralight dark matter.

The coefficients $\gamma_{\ell m_l}$ entering the above sum are purely numerical and are given by\vspace{-0.5em}
\begin{multline}
\gamma_{\ell m_l}=(-1)^{m_l}\frac{(\ell-m_l)!}{(\ell-m_l-2)!} 
\times \\ \left\{\Gamma\!\left(\frac{1-\ell-m_l}{2}\right)
\Gamma\!\left(1+\frac{\ell-m_l}{2}\right)
\Gamma\!\left(\frac{3-\ell+m_l}{2}\right)
\Gamma\!\left(1+\frac{\ell+m_l}{2}\right)\right\}^{-1},
\end{multline}
where $\Gamma(x)$ denotes the Euler gamma function. These coefficients weight the contribution of individual angular modes to the total friction force and ensure the convergence of the multipole expansion.

The functions $S_{\ell,\ell-1}^{m_l}$ appearing in Eq.~\eqref{FDF1_re} have the form\vspace{-0.5em}
\begin{multline}
S_{\ell,\ell-1}^{m_l}= \frac{\pi {\rm i}}{2 \sqrt{1+\frac{m_l^2}{\ell_{\rm q}^2}} }
\bigg[(-1)^{1+\theta(m_l)}j_\ell\!\left(\ell_{\rm q} \mathcal{M} f^-_{m_l}\right)
j_{\ell-1}\!\left(\ell_{\rm q} \mathcal{M} f^-_{m_l}\right)
+\\ {\rm i}\, j_\ell\!\left(\ell_{\rm q} \mathcal{M} f^-_{m_l}\right)
y_{\ell-1}\!\left(\ell_{\rm q} \mathcal{M} f^-_{m_l}\right)
-j_\ell\!\left({\rm i}\ell_{\rm q} \mathcal{M} f^+_{m_l}\right)
h_{\ell-1}^{(1)}\!\left({\rm i}\ell_{\rm q} \mathcal{M} f^+_{m_l}\right)\bigg],
\label{Slm_re}
\end{multline}
where $j_\ell(x)$ and $y_\ell(x)$ are the spherical Bessel functions of the first and second kind, respectively, while $h_\ell^{(1)}(x)$ denotes the spherical Hankel function of the first kind. The Heaviside step function $\theta(x)$ ensures the correct treatment of states with different magnetic quantum numbers.

The auxiliary functions $f_{m_l}^{\pm}$ are defined as\vspace{-0.5em}
\begin{equation}
f^{\pm}_{m_l} = \sqrt{2}\left(\sqrt{1+\left(\frac{m_l}{\ell_{\rm q}}\right)^2}\pm1\right)^{1/2},
\label{fpm_re}
\end{equation}
and depend on the dimensionless parameter $\ell_{\rm q}$,\vspace{-0.5em}
\begin{equation}\label{Lq_re}
\ell_{\rm q}=\frac{r m c_s}{\hbar \mathcal{M}},
\end{equation}
which characterizes the ratio between the macroscopic orbital scale and the microscopic quantum scale associated with the ULDM field.

\begin{figure}
    \centering
    \includegraphics[width=0.9\linewidth]{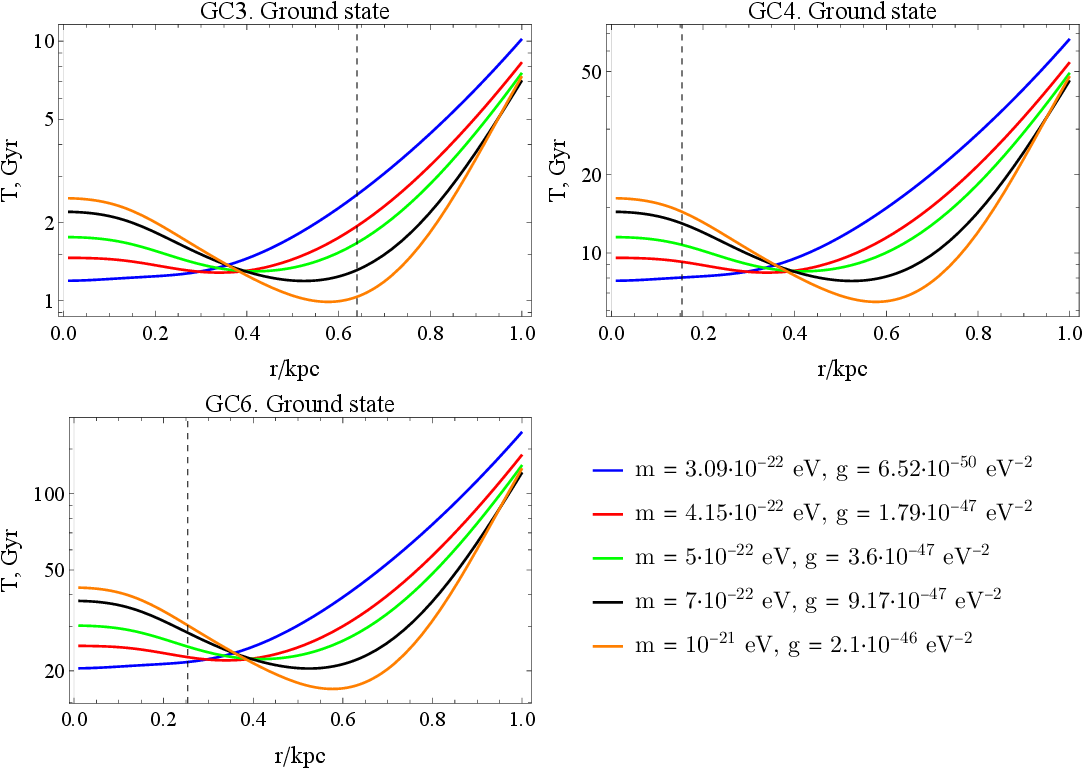}%{Fig1.jpg}%{Ts0.eps}
    \caption{The characteristic time when GC changes notably
its velocity for the spherically symmetric soliton ground state of ULDM at a few values of bosonic particle mass $m$ with the corresponding value of $g$ as a function of its position $r$ from the galactic center. The dotted vertical line shows the current position of a globular cluster.}
    \label{fig:1}
\end{figure}

A key quantity controlling the behavior of the dynamical friction force is the Mach number,
\begin{equation}\label{Mach_re}
\mathcal{M}=\frac{v}{c_s},
\end{equation}
where $v$ denotes the magnitude of the relative velocity between the probe ($v_{GC}$) and the local ULDM flow $\mathbf{u}$.  {Note that the absolute value of the dynamic friction force depends only on the modulus of the relative velocity of the
test particle relative to dark matter, see more details in \cite{Hui:2016ltb}. Hence,} for the vortex ULDM soliton state, there are two possibilities of the direction of probe motion with respect to ULDM:  counter-rotating with the relative velocity $v=v_{GC}+u$ and corotating probe motion when $v=|v_{GC}-u|$.

 {Since, according to the results in Fig.\ref{figure1}, the baryonic matter produces a significant correction to the dark matter gravitational potential, it is necessary to include in the dynamical friction force the contribution due to the baryonic matter. For a probe moving with velocity $\mathbf{v_{GC}}$ in the homogeneous baryonic matter with density $\rho_B$, this contribution is given by the Chandrasekhar dynamical friction force \cite{Chandrasekhar}
\begin{equation}\label{Chander}
   F_{fr, B}
= -\,4\pi G^2 M^2 \rho_B \,
\frac{\ln\Lambda}{v_{GC}^2}
\left[
\operatorname{erf}(X)
- \frac{2X}{\sqrt{\pi}} e^{-X^2}
\right],
\qquad
X \equiv \frac{v_{GC}}{\sqrt{2}\,\sigma}.
\end{equation}
Here $\sigma$ is the velocity dispersion of the background particles (assumed to follow a Maxwellian distribution), which is taken to be $\sigma=10$ km/s \cite{walker2007velocity}. To avoid unphysical results at small distances, 
we follow  \cite{Bar:2022liw} and use the regularized Coulomb logarithm 
\begin{equation}
    \ln \Lambda \rightarrow \frac12\ln(1+\Lambda^2) =\frac12 \ln\left(1+\left(\frac{r\, v_{GC}}{G M}\right)^2\right).
\end{equation}
We use in Eq.\eqref{Chander} the globular cluster velocity $v_{GC}$, because we assume that the averaged velocity of the baryonic matter is zero. }

\begin{figure}[t]
    \centering
    \includegraphics[width=0.9\linewidth]{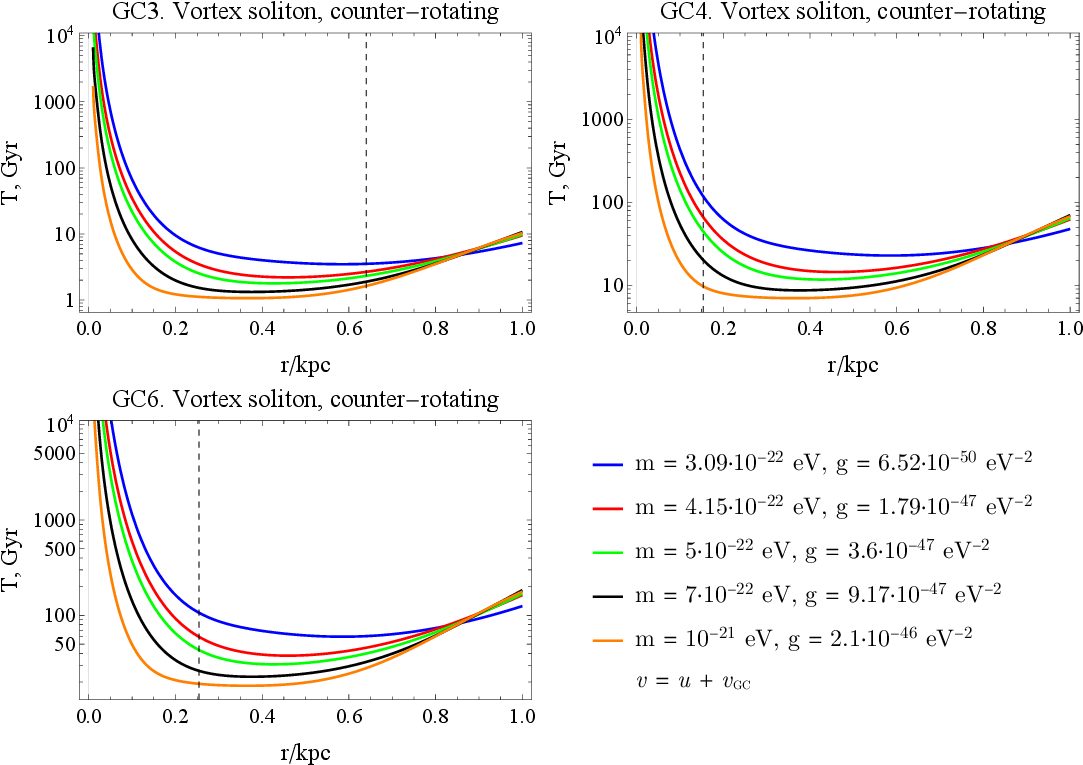}%{Fig2.jpg}%{Ts1d1.eps}
    \caption{The characteristic time when GC changes notably
its velocity for the soliton vortex state of ULDM and a few values of bosonic particle mass $m$ with the corresponding value of $g$ as a function of its position $r$ from the galactic center. The direction of rotation of the BEC and the GC is opposite, and the dotted vertical line shows the current position of a globular cluster.}
    \label{fig:2}
\end{figure}

 {Note that since the acceleration due to the dynamical force is proportional to the probe mass $M$, one would expect the effect of mass segregation, i.e. the tendency of more massive objects to lose kinetic energy through dynamical friction and gradually sink toward the galactic center, while less massive objects remaining at larger distances \cite{BinneyTremaine2008}.}

Further, since the dynamical friction force scales with the square of the mass of the moving object, its impact on globular clusters is expected to be significantly stronger than in the case of individual stars because GCs are much more massive than stars. As a result, globular clusters serve as particularly sensitive probes of dynamical friction effects in dark matter halos. Observationally, the survival of globular clusters at their present-day distances implies that the corresponding characteristic timescale must be of order $\mathcal{O}(10~\mathrm{Gyr})$, comparable to or exceeding the age of the Universe.

Using Eq.~(\ref{probeT}), we computed the characteristic timescales for three globular clusters, specified by their masses and current projected distances from the galactic center: GC3 ($M_{\mathrm{GC}} = 4.98 \times 10^5 M_\odot$, $r_\perp = 0.64$~kpc), GC4 ($M_{\mathrm{GC}} = 0.76 \times 10^5 M_\odot$, $r_\perp = 0.154$~kpc), and GC6 ($M_{\mathrm{GC}} = 0.29 \times 10^5 M_\odot$, $r_\perp = 0.254$~kpc). The analysis was carried out for both the solitonic ground state and the rotating vortex configurations of the ULDM halo, assuming that the clusters are located in the equatorial plane $z = 0$. The calculations were performed using the numerical and analytical techniques developed in \cite{Gorkavenko:2024ocl}.

\begin{figure}
    \centering
    \includegraphics[width=0.85\linewidth]{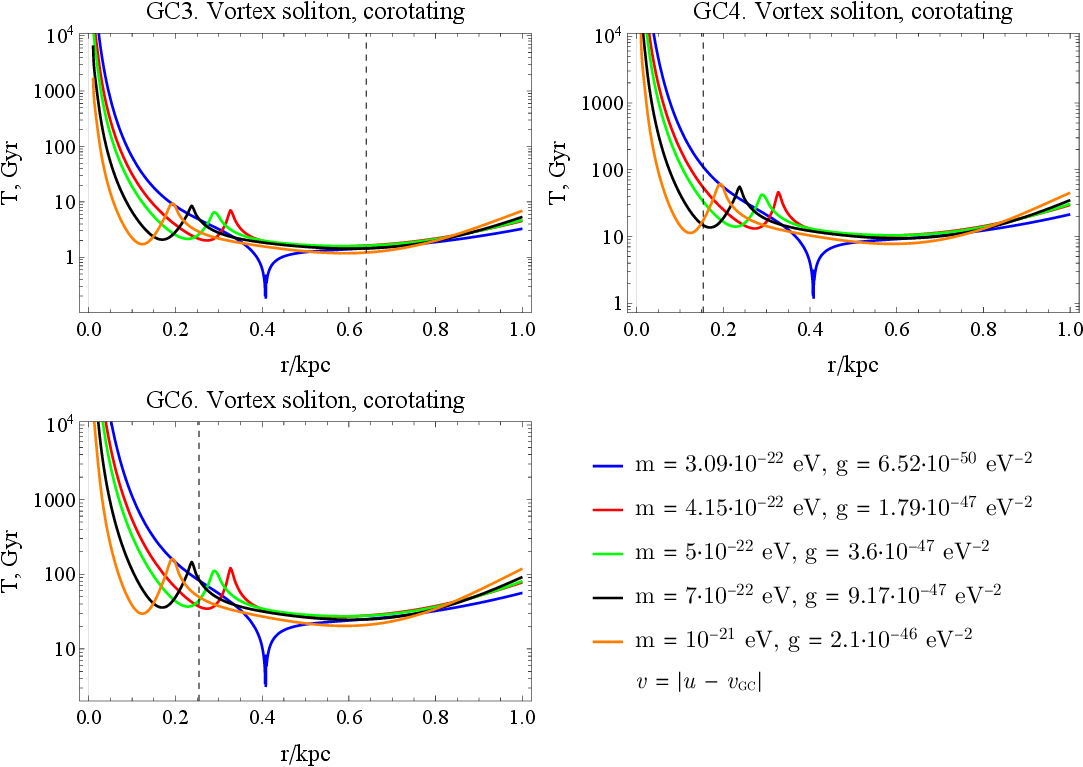}%{Fig3.jpg}%{Ts1d-1.eps}
        \caption{The characteristic time when GC changes notably
its velocity for the soliton vortex state of ULDM and a few values of bosonic particle mass $m$ with the corresponding value of $g$ as a function of its position $r$ from the galactic center. The direction of rotation of the BEC and the GC coincides and the dotted vertical line shows the current position of a globular cluster.}
    \label{fig:3}
\end{figure}

The resulting characteristic times of the velocity substantial change are shown in Figs.~\ref{fig:1}–\ref{fig:3}. A pronounced modification of the globular cluster dynamics is observed in the presence of a vortex state, particularly in the case of co-directional rotation of the cluster and the ULDM flow, where peaks are clearly seen in Fig.\ref{fig:3}.

Analyzing the obtained graphical dependencies, one can see that for both the spherically symmetric stationary soliton model and the vortex soliton model, the characteristic time for a significant change in velocity is of the order of 10 Gyr for globular clusters GC4 and GC6. 

 {The GC3 case requires special consideration. For a soliton in the ground state, one can observe a decreasing behavior of the characteristic timescale at distances smaller than 1 kpc and up to the present-day position of GC3. At small distances from the galactic center, the characteristic timescale is only of a few Gyr.
In the case of a vortex state of the soliton, a very significant increase of the characteristic timescale at small galactocentric distances is observed. This is related to the sharp decrease of the ULDM density inside the vortex soliton, which has a toroidal structure. In the case of corotation between GC3 and the vortex soliton, notable peaks in the characteristic timescale appear at radii where the relative velocity between the ULDM and GC3 vanishes.}

 {Finally, we would like to add that qualitatively, the effect of a rotating structure in ULDM is similar to the standard core stalling picture in cold DM \cite{Petts2015CoreStalling,KaurStone2022DensityWakes,Banik_2021,DiCintioMarcos2025ForceFluctuations}.  As in the case of a cold DM, the dynamical friction force arising in the case of corotation is weakened. This agrees with the classical consideration of dynamical friction by Chandrasekhar \cite{Chandrasekhar}.}

%\newpage
\section{Conclusions}

We have investigated the impact of a rotating vortex state of ultralight dark matter on the dynamical friction acting on globular clusters moving inside a ultralight dark matter core. By directly comparing the resulting dynamical friction force with that obtained for the non-rotating solitonic ground state, we have demonstrated that the internal structure and quantum state of the ULDM core play a crucial role in determining the orbital evolution of massive substructures. Our analysis shows that the wave nature of ULDM, combined with coherent rotation in the vortex configuration, can substantially modify the effective drag force experienced by globular clusters.

One of key results of this work is the finding that co-directional rotation between the vortex flow of the ULDM halo and the orbital motion of a globular cluster leads to a pronounced suppression of dynamical friction  {at some distances leading to the appearance of the peaks in the characteristic time}. This effect arises from the reduction of the relative velocity between the cluster and the surrounding dark matter medium, which directly weakens the gravitational wake responsible for dynamical friction. 

 {
An inspection of the resulting curves shows that, for both the spherically symmetric stationary soliton and the vortex soliton models, the characteristic timescale associated with a substantial change in velocity gives infall times exceeding the age of Universe for the globular clusters GC4 and GC6 at the current position.}

  {The situation is qualitatively different for GC3 and therefore requires separate discussion. In the case of a soliton in the ground state, the characteristic timescale decreases at galactocentric distances below 1 kpc, remaining reduced up to the present-day location of GC3. In the inner regions of the galaxy, this timescale is only a few Gyrs.}

 {For a vortex soliton, by contrast, the characteristic timescale exhibits a pronounced growth at small galactocentric radii. This behavior can be attributed to the strong suppression of the ULDM density in the inner region of the vortex soliton, which has a toroidal density profile. When GC3 corotates with the vortex soliton, distinct peaks in the characteristic time emerge at radii where the relative velocity between the ULDM background and GC3 approaches zero.}

More broadly, these findings highlight the importance of going beyond the ground-state description of ultralight dark matter halos and considering rotating configurations when modeling the dynamics of satellite galaxies and their internal stellar systems. The sensitivity of dynamical friction to the internal velocity structure of ULDM halos suggests that globular cluster dynamics can serve as an indirect probe of the quantum state and coherence properties of dark matter. Future observational constraints on globular cluster orbits, combined with improved modeling of ULDM halo configurations, may therefore provide a novel avenue for testing ultralight dark matter scenarios and distinguishing them from conventional cold dark matter models.

 {The obtained results for the characteristic time of globular clusters indicate the importance of further research on the evolution of globular clusters from a starting point to their current position.}

 {For a vortex soliton state of DM, an interesting question for further investigation is the consideration of a stellar bar embedded
in ULDM halo of the galaxy, because the torque force is significantly different from that in a standard cold
dark matter halo \cite{Weinberg1985BarredGalaxies}. Such consideration can place important constraints on the mass of ULDM particles
forming the halo. }

\section*{Acknowledgments}
K.K. acknowledges funding by the Deutsche Forschungsgemeinschaft (DFG, German Research Foundation) under Germany’s Excellence Strategy– EXC 2123 Quantum Frontiers – 390837967. The work of E.G., V.G., T.G. and A.Z. was partially supported by the project 'Search for dark matter and particles beyond the Standard Model' of the Ministry of Education and Science of Ukraine 25BF051-01. The work of O.T. was partially supported by the Ministry of Education and Science of Ukraine (project №0124U001660).\newpage

\bibliography{bibliography}
\bibliographystyle{JHEP}

\end{document}